\documentclass[conference]{IEEEtran}
\usepackage[dvipdfm]{graphicx}
\usepackage[usenames,dvipsnames]{color}
\usepackage{amsmath}%
\usepackage{amsfonts}%
\usepackage{amssymb}%
\usepackage{graphicx}%
\usepackage{multicol}%
\usepackage{verbatim}%
\usepackage{enumerate}%

\begin{document}

\title{Cognitive Power Control Under Correlated Fading and Primary-Link CSI}



\author{
\authorblockN{Doha Hamza, Mohammed Nafie}
\authorblockA{Wireless Intelligent Networks
Center (WINC)\\
Nile University, Cairo, Egypt\\
Email: doha.hamza@nileu.edu.eg, mnafie@nileuniversity.edu.eg}}

\maketitle

\begin{abstract}

We consider the cognitive power control problem of maximizing the secondary throughput under an outage probability constraint on a constant-power constant-rate primary link. We assume a temporally correlated primary channel with two types of feedback: perfect delayed channel state information (CSI) and one-bit automatic repeat request (ARQ). We use channel correlation to enhance the primary and secondary throughput via exploiting the CSI feedback to predict the future primary channel gain. We provide a numerical solution for the power control optimization problem under delayed CSI. In order to make the solution tractable under ARQ-CSI, we re-formulate the cognitive power control problem as the maximization of the instantaneous weighted sum of primary and secondary throughput. We propose a greedy ARQ-CSI algorithm that is shown to achieve an average throughput comparable to that attained under the delayed-CSI algorithm, which we solve optimally. 

\end{abstract}

\section{Introduction}
\label{sec:Introduction}

Cognitive technology is bound to change the way spectrum is accessed and used. Cognitive power control with the conflicting goals of maximizing secondary users (SUs) throughput and minimizing interference to primary users (PUs) is a key enabling technology for cognitive devices. 

Feedback-based cognitive power control recently received attention in the cognitive literature (e.g., \cite{Zhang}, \cite{zhang3}, \cite{fabio}, \cite{bits}).  In \cite{bits}, for instance, the authors suggest an on-off SU power control scheme based on observing the automatic repeat request (ARQ) feedback from primary receiver which reflects PU achieved packet rate. The key difference between our work and others is that we assume a temporally correlated channel between the primary transmitter and receiver. Based on primary channel state information (CSI) feedback, a cognitive transmitter can exploit the correlation to predict the primary link  gain during the next transmission phase.\footnote{In using a correlated channel model for cognitive power control with feedback, we are motivated by existing rate adaptation literature that assumes existence of the Markovian property (e.g., \cite{R.Aggarwal} and \cite{linnartz}).} The cognitive power is adjusted accordingly so that the secondary throughput is maximized and the primary link outage probability is kept below a certain maximum. In section \ref{sec:sim_results} we compare the performance of our algorithm with the one proposed in \cite{bits} where no knowledge of the channel statistics is assumed. We show via simulations that, under channel correlation assumptions, our model garners better performance out of the ARQs.

The system model and assumptions are described in Section \ref{sec:SysMdl}. In Section \ref{sec:const_optim}, we pose the cognitive power optimization problem and solve it optimally under delayed CSI. We formulate the cognitive power control problem as a maximization of the weighted sum of the primary and secondary throughput under various CSI in Section \ref{sec:outage_prob}.  Section \ref{sec:sim_results} presents the numerical results. Our work is concluded in Section \ref{sec:Conc}.

\section{System Model}
\label{sec:SysMdl}
\begin{figure}
	\centering
		\includegraphics[scale=0.70]{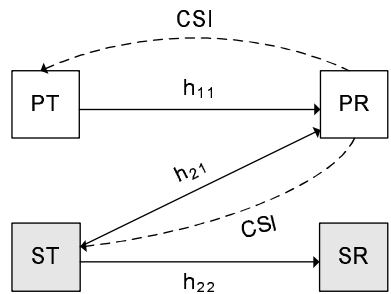}
	\caption{Cognitive Power Control via ARQ: System model}
	\label{fig:mdl}
\end{figure}

We consider a pair of primary and secondary users as depicted in Fig. \ref{fig:mdl}. The secondary pair is composed of the secondary transmitter (ST) and secondary receiver (SR). The secondary pair attempts to share the spectrum resource with the primary link, composed of the primary transmitter (PT) and primary receiver (PR). The goal of the secondary terminals is to maximize their throughput, via the choice of ST transmit power, provided that the primary system outage probability is kept below a maximum value, $P_{out}$. 

The PT transmits all the time with a constant power $p^{p}$, and a constant rate $R_o$, both known to the secondary transmitter. We ignore the interference from PT on SR.\footnote{This is a valid assumption under a short-range cognitive channel or when SR employs a technique for canceling the interference from PT.} We also assume that $h_{22}$ and $h_{21}$ are known to the secondary transmitter, and constant. Hence, $h_{22}(t)=h_{22}$ and  $h_{21}(t)=h_{21}$, where t is the time index. 

The ST overhears the primary link-layer CSI feedback from PR to PT, assumed to be transmitted over an error-free control channel. We consider two types of CSI availability in this paper: delayed and 1-bit ARQ-CSI. In delayed CSI feedback, PR is assumed to send to PT a perfect unquantized estimate of the primary channel gain experienced by the last packet. Hence, we also refer to this type of transmitted feedback as causal genie CG-CSI.

The primary link channel follows a block-fading Markov channel model where the channel complex gain varies from packet to packet according to the following first order autoregressive equation:

\begin{equation}
h_{11}(t)={(1-\alpha)}h_{11}(t-1)+\sqrt{2\alpha-\alpha^{2}}w(t)
\label{eq:primary_channel}
\end{equation}

\noindent where $w(t)$ is a zero-mean unit-variance circularly symmetric complex Gaussian innovation process, and $\alpha\in[0,1]$ determines the channel correlation. Note that $\alpha=1$ corresponds to i.i.d. gains, and  $\alpha=0$ corresponds to a time invariant gain. The channel gain at $t=0$, $h_{11}\left(0\right)$, is independent of $w(t)$ and is also  
a zero-mean unit-variance circularly symmetric complex Gaussian process. This makes the process $h_{11}\left(t\right)$ stationary with a unit-variance Rayleigh-distributed magnitude.  

In the sequel, we let $g_{21}=|h_{21}|^{2}$, $g_{22}=|h_{22}|^{2}$, and $\sigma_p^2 $ and $\sigma_s^2 $ are the noise variances at the primary and secondary receivers, respectively. To distinguish the primary link channel gain over which the CSI is fed back, we let $\gamma_{t}=|h_{11}(t)|^{2}$. For the CG-CSI, $\gamma_{t-1}$ is made available. We denote the ARQ-CSI as $(\hat{a}_{t-1})=(\hat{\epsilon} _{t-1},\hat{p}^{s}(\hat{a}_{t-2}))$, where the hat symbol denotes vectors, $\hat{\epsilon} _{t-1}=[\epsilon_{1}\quad \epsilon_{2}...\epsilon_{t-1}]$ is a vector of ARQ-CSI feedback and the secondary power policy $\hat{p}^{s}(\hat{a}_{t-2})=[p^{s}(\hat{a}_{0})\quad p^{s}(\hat{a}_{1})...p^{s}(\hat{a}_{t-2})]$ is a vector of previous ST transmit power decisions. For the ARQ-CSI, we assume an explicit feedback model where $\epsilon_{t}$ for packet $t$ is $\epsilon_{t}=1$ for correct reception, and $\epsilon_{t}=0$ for erroneous reception.\footnote {In current extension of this work, we consider an $h_{21}{(t)}$ channel update model similar to the primary link model used in (\ref{eq:primary_channel}) and include a training phase to learn $h_{21}$.}

\section{The cognitive Power Control Problem as a Constrained Optimization Problem}
\label{sec:const_optim} 

The cognitive power control problem can be posed as a constrained optimization problem as follows:

\begin{equation}
\begin{split}
\max.\quad&\lim_{L\rightarrow\infty}\frac{1}{L}\sum^{L}_{i=1}\left[\mathbb{E}_{\hat{d}_{i-1}}{\log\left(1+\frac{p^{s}(\hat{d}_{i-1})g_{22}}{\sigma_{s}^{2}}\right)}\right]\\
\quad\\
s.t.\quad&\lim_{L\rightarrow\infty}\frac{1}{L}\sum^{L}_{i=1}\left[\mathbb{E}_{\hat{d}_{i-1}}{\int_{0}^{\gamma^{th}}{f(\gamma_{i}|(\hat{d}_{i-1}))d\gamma_{i}}}\right]\leq P_{out},\\
\quad\\
&0\leq p^{s}(\hat{d}_{i-1})\leq p_{max}
\label{eq:constrained}
\end{split}
\end{equation}

\noindent where $\hat{d}_{i-1}$ is a generic vector of CSI available at time $i$ (e.g., $\hat{d}_{i-1}=\hat{a}_{i-1}$ under ARQ-CSI), and $p^{s}(\hat{d}_{i-1})$ is the secondary power value at time $i$ given the CSI at time $(i-1)$. The primary link experiences outage when $\gamma_{i}\leq\gamma^{th}$. Outage is assumed to occur when $R_{o}$ exceeds the primary link capacity. Assuming Gaussian signaling, it is straightforward to obtain $\gamma^{th}$ as: 

\begin{equation}
\gamma^{th}=\left(\exp\left(R_o\right)-1\right)\left(\frac{p^{s}(\hat{d}_{i-1})g_{21}+\sigma_p^{2}}{p^{p}}\right)
\label{eq:threshold_gamma}
\end{equation} 

The objective above denotes the secondary link expected throughput. The first constraint is the expected outage probability of the primary link, whereas the second constraint sets a lower bound of zero on $p^{s}(\hat{d}_{i-1})$ and an upper bound of $p_{max}$. The expectation operator is over the available CSI at time $i$, $\hat{d}_{i-1}$. It is required to find the optimal secondary power vector $\hat{p}^{s*}(\hat{d}_{L-1})=[p^{s*}(\hat{d}_{0}),p^{s*}(\hat{d}_{1}),...,p^{s*}(\hat{d}_{L-1})]$ that maximizes (\ref{eq:constrained}), where the asterisk denotes optimality. We now consider the solution of this problem under the two types of available CSI.

\subsection{The constrained Optimization Problem Under CG-CSI}
\label{sec:cg_constrained} 

Under CG-CSI, solving (\ref{eq:constrained}) is equivalent to solving:

\begin{equation}
\begin{split}
\max. \quad & \mathbb{E}_{\gamma_{t-1}} \left[\log\left(1+\frac{p^{s}(\gamma_{t-1})g_{22}}{\sigma_{s}^{2}}\right)\right]\\
\quad\\
s.t.\quad & \mathbb{E}_{\gamma_{t-1}}\left[\int_{0}^{\gamma^{th}}{f(\gamma_{t}|\gamma_{t-1})d\gamma_{t}}\right]\leq P_{out},\\
\quad\\
& 0\leq p^{s}(\gamma_{t-1})\leq p_{max}
\label{eq:constrained_cg}
\end{split}
\end{equation}

\noindent where $\gamma^{th}$ is evaluated using (\ref{eq:threshold_gamma}) with $\hat{d}_{t-1}=\gamma_{t-1}$. The summation in (\ref{eq:constrained}) is eliminated in (\ref{eq:constrained_cg}) because the optimal power control policy is greedy under CG-CSI \cite{R.Aggarwal}. Since the optimal future power assignments $\left\{p^{s*}(\gamma_{k-1})\right\}_{k > t}$ are chosen based on perfect knowledge of delayed channel gains, they do not depend on $p^{s}(\gamma_{t-1})$ rendering optimal the greedy formulation in (\ref{eq:constrained_cg}). Note that (\ref{eq:constrained_cg}) is a non-convex problem to which we present a numerical solution in Section \ref{sec:sim_results}. 

\subsection{The constrained Optimization Problem Under ARQ-CSI}
\label{sec:arq_constrained}
In the case of ARQ-CSI, the optimal cognitive power control assignment at a given instant conditioned on the degraded feedback is a partially observable Markov decision process (POMDP) \cite{R.Aggarwal}. Since the transmitted power at a given instant influences the ARQ-CSI feedback received in the future, the optimal power at any instant depends on the optimal power at subsequent instances. Hence, solving (\ref{eq:constrained}) is intractable under ARQ-CSI. In order to reduce the computational complexity of the solution in the case of ARQ-CSI, we present a reformulation of the constrained optimization problem in the next section.  This new formulation makes the ARQ-CSI power control problem amenable to online implementation.

\section{The Cognitive Power Control Problem as the Weighted Sum Throughput}
\label{sec:outage_prob}

In this section, we pose the cognitive power control problem as the maximization of the weighted sum of primary and secondary throughput. Moreover, we replace the expectation over CSI with time averaging\footnote{This is valid for ergodic Markov chains since the long-term proportion of time spent in a given state approaches the state's steady state probability  \cite{garcia}.} and consider the maximization of the instantaneous weighted sum throughput.

\subsection{Weighted Sum Throughput Under CG-CSI}
\label{sec:cg} 

For illustration purposes, we first apply the weighted sum formulation to the relatively-easier case of CG-CSI which is later used to upper-bound the performance under ARQ-CSI. The instantaneous weighted sum throughput $T(p^{s}(\gamma_{t-1}))$ is given by:

\begin{equation}
\begin{split}
T(p^{s}(\gamma_{t-1}))=&\left({1-\beta}\right)\log\left(1+\frac{p^{s}(\gamma_{t-1})g_{22}}{\sigma_{s}^{2}}\right)+\\
&\beta{R_{o}}\left[1-\int_{0}^{\gamma^{th}}{f(\gamma_{t}|\gamma_{t-1})d\gamma_{t}}\right]
\label{eq:thru}
\end{split}
\end{equation}

\noindent where $\beta\in\left[0,1\right]$ is a weighting parameter that reflects the relative importance of the primary and secondary throughput. With the proper choice of $\beta$, maximizing the weighted sum implicitly enforces the target outage probability for the primary link \cite{raphael}. The goal is to find the optimal $p^{s*}(\gamma_{t-1})$ value that maximizes (\ref{eq:thru}), where $p^{s}(\gamma_{t-1})\in\left[0,p_{max}\right]$. We calculate the average throughput as: 

\begin{equation}
T_{\rm CG}^{av}=\frac{1}{K}\left\{\sum^{K}_{k=1}{T(p^{s*}(\gamma_{k-1}))}\right\}
\label{eq:csi_arq3}
\end{equation}

\noindent where K is the number of instants over which the $T_{\rm CSI}^{av}$ is calculated.

The outage probability varies according to: (a) the value of the primary link channel gain experienced by a packet and (b) the secondary power level $p^{s}(\gamma_{t-1})$. From the channel model given in (\ref{eq:primary_channel}), it can be shown \cite{R.Aggarwal} that the pdf of $\gamma_{t}$ conditioned on the delayed-CSI is:

\begin{equation}
\begin{split}
f{(\gamma_{t}|{\gamma_{t-1}})}& ={\frac{1}{2\alpha-\alpha^2}}\exp({-\frac{\gamma_{t}+({1-\alpha})^{2}\gamma_{t-1}}{2\alpha-\alpha^2}})\times\\
& \quad \\
& \quad I_{o}\left(\frac{2(1-\alpha)}{2\alpha-\alpha^2}\sqrt{\gamma_{t}\gamma_{t-1}}\right)
\label{eq:recursive}
\end {split}
\end{equation}

\noindent where $I_{o}(.)$ is the first order modified Bessel function.

We now summarize the steps used to calculate the average sum throughput under CG-CSI:

\begin{enumerate}
\item Using (\ref{eq:recursive}), calculate the conditional outage probability corresponding to a certain $\gamma_{k-1}$. 
\item At each time instant, find the optimal $p^{s*}(\gamma_{k-1})$ value that maximizes the instantaneous weighted sum throughput using (\ref{eq:thru}).
\item The average throughput $T_{\rm CG}^{av}$ is obtained from (\ref{eq:csi_arq3}).
\end{enumerate}

The maximization of the weighted sum is a scalarization technique of the constrained optimization problem described in the previous section. The optimal pair $(p^{s*}(\gamma_{t-1}),\beta)$ obtained from (\ref{eq:thru}) are Pareto-optimal solutions for the constrained optimization problem. However, for some non-convex problems, the scalarization technique does not return all the Pareto-optimal points \cite{boyd}. In other words, solving the constrained optimization problem is more general than solving (\ref{eq:thru}).

\subsection{Weighted Sum Throughput Under ARQ-CSI}
\label{sec:arq}

In this section we would like to replace the non-degraded CG-CSI with the 1-bit degraded ARQ-CSI feedback. As described previously, the optimal ACK-CSI policy is a POMDP process. Therefore, when the instantaneous throughput is used, the optimal power $p^{s*}(\hat{a}_{t-1})$ is given by:

\begin{equation}
\begin{split}
p^{s*}(\hat{a}_{t-1})=\arg\max_{p^{s}(\hat{a}_{t-1})}\Big{[}&T(p^{s}(\hat{a}_{t-1}))+\\&\sum^{T}_{k=t+1}{T(p^{s*}(\hat{a}_{k-1}))}\Big{]}
\label{eq:csi}
\end{split}
\end{equation}

\noindent where $p^{s}(\hat{a}_{t-1})\in\left[0,p_{max}\right]$, and $T(p^{s}(\hat{a}_{t-1}))$ is evaluated as in (\ref{eq:thru}) using the ARQ-CSI. Note that (\ref{eq:csi}) is prohibitive to solve. Inspired by the optimal CG-CSI scheme, we implement a greedy scheme to maximize the weighted sum throughput as follows:

\begin{equation}
p^{s}(\hat{a}_{t-1})=\arg\max_{p^{s}(\hat{a}_{t-1})}\left\{{T(p^{s}(\hat{a}_{t-1}))}\right\}
\label{eq:csi_arq}
\end{equation}

\noindent note that we now removed the asterisk with the implementation of the greedy scheme .The average throughput is obtained from (\ref{eq:csi_arq3}) using the ARQ-CSI, where we replace $p^{s*}(\hat{a}_{t-1})$ with the outcome of the greedy ARQ algorithm, i.e. $p^{s}(\hat{a}_{t-1})$. To calculate the outage probability, we derive the pdf $f(\gamma_{t}|\hat{a}_{t-1},p^{s}(\hat{a}_{t-1}))$. Following a similar derivation to \cite{R.Aggarwal}, we write:
\begin{align}
\label{eq:original}
f(\gamma_{t}|\hat{a}_{t-1}, p^{s}(\hat{a}_{t-1}))&=f(\gamma_{t}|\hat{\epsilon}_{t-1}, \hat{p}^{s}(\hat{a}_{t-1}))\\\nonumber
\quad\quad\quad\quad &= \int f\left(\gamma_{t}|\gamma_{t-1},\hat{\epsilon}_{t-1},\hat{p}^{s}(\hat{a}_{t-1})\right)\times\\\nonumber
& \quad\quad f\left(\gamma_{t-1}| \hat{\epsilon}_{t-1},\hat{p}^{s}(\hat{a}_{t-1})\right)d{\gamma_{t-1}}\\\nonumber
&= \int f(\gamma_{t}|\gamma_{t-1})\times\\\nonumber
&\quad\quad f{(\gamma_{t-1}| {\hat{\epsilon}_{t-1},\hat{p}^{s}(\hat{a}_{t-1})})}d{\gamma_{t-1}}
\end{align}

\noindent where in (\ref{eq:original}), we make use of the Markovian property. Note that we also restored $\hat{a}_{t-1}$ to its basic definition in Section \ref{sec:SysMdl}. Using (\ref{eq:recursive}), we re-write the second part of equation (\ref{eq:original}) using Bayes' rule as:

\begin{align}
&f{(\gamma_{t-1}|{\hat{\epsilon}_{t-1},\hat{p}^{s}(\hat{a}_{t-1})})}=f{(\gamma_{t-1}|\epsilon_{t-1},{\hat{\epsilon}_{t-2},\hat{p}^{s}(\hat{a}_{t-1})})}= \nonumber \\
&\quad \nonumber \\  &\frac{\Pr({\epsilon_{t-1}|\gamma_{t-1},\hat{\epsilon}_{t-2},\hat{p}^{s}(\hat{a}_{t-1})})f({\gamma_{t-1}|\hat{\epsilon}_{t-2},\hat{p}^{s}(\hat{a}_{t-1})})}{\int \Pr({\epsilon_{t-1}|\gamma_{t-1},\hat{\epsilon}_{t-2},\hat{p}^{s}(\hat{a}_{t-1})})f({\gamma_{t-1}|\hat{\epsilon}_{t-2},\hat{p}^{s}(\hat{a}_{t-1})})d{\gamma_{t-1}}}
\label{eq:bayes}
\end{align}

Note that,
\begin{equation}
\Pr({\epsilon_{t-1}|\gamma_{t-1},\hat{\epsilon}_{t-2},\hat{p}^{s}(\hat{a}_{t-1})})=\Pr({\epsilon_{t-1}|\gamma_{t-1},p^{s}(\hat{a}_{t-2})})
\label{eq:bayessplit}
\end{equation}

\noindent Equation (\ref{eq:bayessplit}) follows from the fact that the ACK/NACK for a given time instant $(t-1)$ is only a function of the primary channel gain at time $(t-1)$ and the secondary transmitter power at $(t-1)$, $p^{s}(\hat{a}_{t-2})$. Based on our assumption that error occurs when the transmission rate exceeds the capacity, we get the following values for the conditional probability ${\rm Pr}({\epsilon_{t-1}|\gamma_{t-1},p^{s}(\hat{a}_{t-2})})$:

\begin{equation}
{\rm Pr}({\epsilon_{t-1}=1|\gamma_{t-1},p^{s}(\hat{a}_{t-2})})=\left\{
\begin{array}{l l}
1,\quad\quad\mbox{$ \gamma_{t-1}\geq \gamma^{th}$}\\
0,\quad\quad \mbox{$else$}
\label{eq:epsilon_pdf1}
\end{array}
\right.
\end{equation}

\begin{equation}
{\rm Pr}({\epsilon_{t-1}=0|\gamma_{t-1},p^{s}(\hat{a}_{t-2})})= \left\{
\begin{array}{l l}
1,\quad \quad  \mbox{$ \gamma_{t-1}\leq \gamma^{th}$}\\
0,\quad \quad \mbox{$else$}
\label{eq:epsilon_pdf2}
\end{array}
\right.
\end{equation}

\noindent where $\gamma^{th}$ is evaluated using (\ref{eq:threshold_gamma}) with $\hat{d}_{t-2}=\hat{a}_{t-2}$. Note also in (\ref{eq:bayes}) that:

\begin{equation}
f{(\gamma_{t-1}|{p^{s}(\hat{a}_{t-1}),\hat{\epsilon}_{t-2},\hat{p}^{s}(\hat{a}_{t-2})})}=f{(\gamma_{t-1}|{\hat{\epsilon}_{t-2},\hat{p}^{s}(\hat{a}_{t-2})})}
\label{eq:bayes2}
\end{equation}

\noindent Hence, (\ref{eq:bayes}) becomes:

\begin{align}
\label{eq:bayes1_final}
&{f{(\gamma_{t-1}|\hat{\epsilon}_{t-1},\hat{p}^{s}(\hat{a}_{t-1}))}}=\\\nonumber
&\quad \\\nonumber
&{\frac{\Pr(\epsilon_{t-1}|\gamma_{t-1},p^{s}(\hat{a}_{t-2}))f({\gamma_{t-1}|\hat{\epsilon}_{t-2},\hat{p}^{s}(\hat{a}_{t-2})})}{\int \Pr({\epsilon_{t-1}|\gamma_{t-1},p^{s}(\hat{a}_{t-2})})f({\gamma_{t-1}|\hat{\epsilon}_{t-2},\hat{p}^{s}(\hat{a}_{t-2})})d{\gamma_{t-1}}}}\\\nonumber
\end{align}

\noindent By incorporating (\ref{eq:bayes1_final}) in (\ref{eq:original}), we have reached a recursive formula for the calculation of the primary channel gain given the history of past ARQ and the secondary transmitter power decisions. 

We now summarize the steps for obtaining the greedy $T_{\rm ARQ}^{av}$:
\begin{enumerate}
  \item Start from a randomly generated primary channel realization.
	\item Initialize the pdf $f{( \gamma_{t-1}| \hat{\epsilon}_{t-2},\hat{p}^{s}(\hat{a}_{t-1}))}$ with the exponential prior.
	\item Calculate the pdf $f({\gamma_{t}|\gamma_{t-1}})$ from equation (\ref{eq:recursive}).
	\item Once a packet is transmitted, and the acknowledgment is received, calculate ${\rm Pr}{(\epsilon_{t-1}|\gamma_{t-1},p^{s}(\hat{a}_{t-2}))}$. 
	\item Obtain $f{(\gamma_{t}| \hat{\epsilon}_{t-1},\hat{p}^{s}(\hat{a}_{t-1}))}$ from (\ref{eq:bayes1_final}) and use it in (\ref{eq:original}) to update $f(\gamma_{t}|\hat{\epsilon}_{t-1},\hat{p}^{s}(\hat{a}_{t-1}))$.
	\item Use $f(\gamma_{t}|\hat{\epsilon}_{t-1},\hat{p}^{s}(\hat{a}_{t-1}))$ to find the $p^{s}(\hat{a}_{t-1})$ value that maximizes (\ref{eq:csi_arq}). Save the throughput attained under this $p^{s}(\hat{a}_{t-1})$ value, $T(p^{s}(\hat{a}_{t-1}))$. 
	\item Update the channel gain using (\ref{eq:primary_channel}) and repeat from step 4. 
\end{enumerate}

The average ARQ-CSI throughput, $T_{\rm ARQ}^{av}$ is obtained by averaging the achieved throughput over time, using (\ref{eq:csi_arq3}).  A fundamental question remains as to how close is this greedy approach to the optimal ACK power control algorithm, $T(p^{s*}(\hat{a}_{t-1}))$. Following a similar argument to \cite{linnartz}, a policy given CG-CSI can always choose not to use this additional CSI and yet achieve optimal ACK-CSI throughput $T(p^{s*}(\hat{a}_{t-1}))$. Using the CG-CSI optimally can only achieve the same or greater throughput, hence $T(p^{s*}(\hat{a}_{t-1})) \leq T(p^{s*}(\gamma_{t-1}))$. In Section \ref{sec:sim_results}, we show via simulations that the average throughput attained greedily under the ARQ-CSI is close to that attained under CG-CSI, and by the reasoning above, the optimal ARQ-CSI policy lies somewhere in between. 

\section{Simulation Results}
\label{sec:sim_results}

For the obtained simulation results, the system parameters are as follows: The rate $R_o$ is chosen as the capacity of an AWGN channel with ${\rm SINR}=10$ dB, {\it i.e.} $R_o=\log(11)\approx2.4$, where $\log\left(.\right)$ denotes the natural logarithm. We assume noise at PR and SR to have unit variance. The primary link's constant transmit power $p^{p}=95$ and the maximum secondary transmit power $p_{max}=20$. Also $g_{21}=1$ and $g_{22}=2$.  We consider $\gamma_{max}=8$.

\begin{figure}
		\includegraphics[scale=0.6]{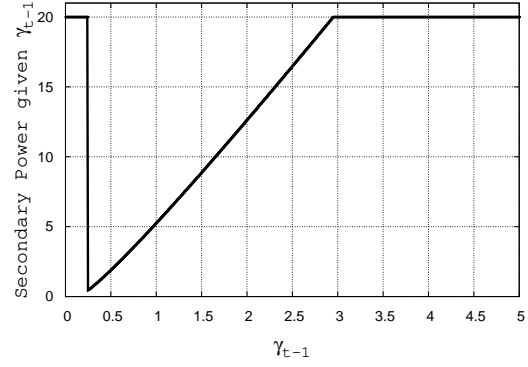}
	\caption{Secondary Power Policy}
	\label{fig:cg_optimization}
\end{figure}

Fig. \ref{fig:cg_optimization} shows the solution of the constrained optimization problem in (\ref{eq:constrained_cg}) using numerical methods, for $\alpha=0.1$, and $P_{out}=25\%$. The result is rather intuitive. At extremely small values of $\gamma_{t-1}$, the primary link is with high probability in natural outage, hence ST uses maximum power, while as the $\gamma_{t-1}$ value increases, ST adjusts the power accordingly to make use of the capacity gap. At large values of $\gamma_{t-1}$, albeit occuring with small probability, ST transmits with maximum power without fear of putting the primary link in outage. The expected throughput attained by averaging over the exponential distribution of $\gamma_{t-1}$ is verified to be equal to the throughput attained via the weighted sum throughput maximization under delayed-CSI.

For the maximization of the weighted sum throughput problem, the simulations are carried out for $\alpha=0.01,0.02,0.05,0.1,\rm and$ $1$. For each $\alpha$ value, we consider 20 $\beta$ values uniformly spaced over the interval $\left[0,0.99\right]$. We average the throughput obtained over 100 random channel realizations for each $\beta$ value. For any channel realization, we send a specified number of packets consistent with the respective $\alpha$-value. Note from (\ref{eq:primary_channel}), that to obtain a correlation factor of $10^{-5}$ between the initial packet and the final one, we need a number of packets$=\frac{-5\log10}{\log(1-\alpha)}$.  For a Rayleigh fading channel with an average power of unity, the primary channel gain, $\gamma_{t}$, is exponentially distributed. We use this pdf in the case where no CSI is available to calculate the outage probability. Note that the exponential distribution also corresponds to the case of $\alpha=1$ as evident from (\ref{eq:recursive}).

\begin{figure}
		\includegraphics[scale=0.65]{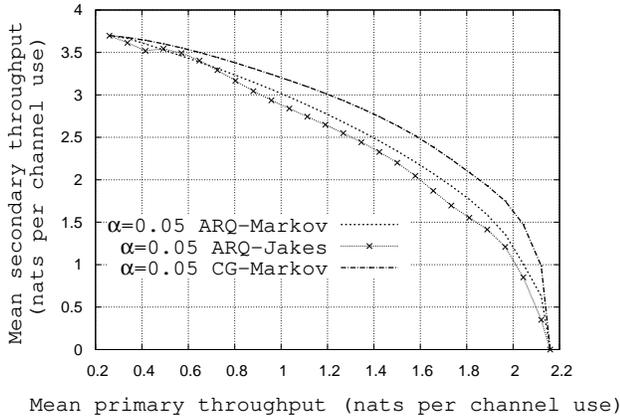}
	\caption{Throughput using the ARQ-CSI and CG-CSI}
	\label{fig:arq_cg}
\end{figure}

Fig. \ref{fig:arq_cg} presents the trade off between the primary and secondary throughputs attained under the ARQ and CG policies at $\alpha=0.05$. It is clear from the figure that the ARQ algorithm is close to the optimal performance, now upper bounded by the causal genie. Note that at $\beta=0$, the primary throughput does not drop to zero, despite the secondary transmit power being always equal to $p_{max}$. This happens as there are possibly large $\gamma$ values that permit the primary link to achieve some throughput despite strong ST interference. At the peak value of $\beta=0.99$, we never attain the $2.4$ value of $R_{o}$. Instead there is almost a $10$ percent loss in the primary throughput, which occurs due to the natural outage of the primary link, i.e. outage not caused by SU, when $R_{o}>\log\left(1+\frac{p^{p}\gamma_{t}}{\sigma_{p}^{2}}\right)$. 

For analysis purposes and algorithmic design, we also superimposed on Fig. \ref{fig:arq_cg} plots of the sum throughput using the 1st order Markov channel model and a real Rayleigh fading channel model, Jakes model. Under low correlation, e.g. $\alpha=0.1$, the two models are the same. For higher values, $\alpha=0.05$, there is a slight channel mismatch evident in Fig. \ref{fig:arq_cg}, which may be mediated by a back-off factor in the calculation of the secondary throughput under the Markov channel model. 

\begin{figure}
		\includegraphics[scale=0.65]{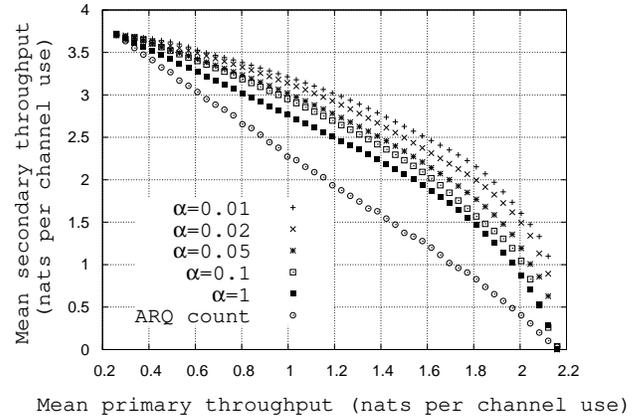}
	\caption{Primary and secondary throughput using the ARQ-CSI, under various $\alpha$ values}
	\label{fig:arq_throughput}
\end{figure}

Fig. \ref{fig:arq_throughput} presents the primary and secondary throughput trade off for different $\alpha$ values. Note that as the correlation between successive channel realizations increases, that the ARQ-CSI algorithm is able to garner better throughput from the system. Note also the increased throughput achieved via ARQ-CSI versus the ARQ-count algorithm presented in \cite{bits}. The case $\alpha=1$, which corresponds to the exponential prior case yields better results than the ARQ-count.

\section{Conclusion}
\label{sec:Conc}
In this paper, we considered feedback-based cognitive power control. Under a block-fading Markov channel model, we posed the cognitive power control problem as a constrained optimization problem, which we solved numerically under delayed-CSI. By re-formulating the problem as the maximization of the weighted sum of primary and secondary throughput, we showed that the 1-bit degraded greedy ARQ-CSI power control algorithm achieves close performance to the optimal CG-CSI policy.  

In future work, we would like to explore the impact of incorporating learning of the channel gain $h_{21}$ in the ARQ-CSI model under a channel update model similar to the one implemented for the primary link gain, $h_{11}$. We would also like to consider a more complicated feedback model, for example where implicit ACK/NACKs are implemented.

\end{document}